# 4MOST Consortium Survey 2: The Milky Way Halo High-Resolution Survey

Norbert Christlieb[1]
Chiara Battistini[1]
Piercarlo Bonifacio[2]
Elisabetta Caffau[2]
Hans-Günter Ludwig[1]
Martin Asplund[3]
Paul Barklem[4]
Maria Bergemann[5]
Ross Church[6]
Sofia Feltzing[6]
Dominic Ford[6]
Eva K. Grebel[7]
Camilla Juul Hansen[5]
Amina Helmi[8]
Georges Kordopatis[9]
Mikhail Kovalev[5]
Andreas Korn[4]
Karin Lind[5]
Andreas Quirrenbach[1]
Jan Rybizki[5]
Ása Skúladóttir[5]
Else Starkenburg[10]

[1] Zentrum für Astronomie der Universität Heidelberg/Landessternwarte, Germany
[2] GEPI, Observatoire de Paris, Université PSL, CNRS, France
[3] Research School of Astronomy & Astrophysics, Australian National University, Canberra, Australia
[4] Department of Physics and Astronomy, Uppsala universitet, Sweden
[5] Max-Planck-Institut für Astronomie, Heidelberg, Germany
[6] Lund Observatory, Lund University, Sweden
[7] Zentrum für Astronomie der Universität Heidelberg/Astronomisches Rechen-Institut, Germany
[8] Kapteyn Instituut, Rijksuniversiteit Groningen, the Netherlands
[9] Observatoire de la Côte d'Azur, Nice, France
[10] Leibniz-Institut für Astrophysik Potsdam (AIP), Germany

We will study the formation history of the Milky Way, and the earliest phases of its chemical enrichment, with a sample of more than 1.5 million stars at high galactic latitude. Elemental abundances of up to 20 elements with a precision of better than 0.2 dex will be derived for these stars. The sample will include members of kinematically coherent substructures, which we will associate with their possible birthplaces by means of their abundance signatures and kinematics, allowing us to test models of galaxy formation. Our target catalogue is also expected to contain 30 000 stars at a metallicity of less than one hundredth that of the Sun. This sample will therefore be almost a factor of 100 larger than currently existing samples of metal-poor stars for which precise elemental abundances are available (determined from high-resolution spectroscopy), enabling us to study the early chemical evolution of the Milky Way in unprecedented detail.

## Scientific context

Galaxy formation simulations predict that the halo of the Milky Way consists in part of stars that formed *in situ* and in part of stars that were accreted from smaller galaxies (for example, Pillepich et al., 2015). The simulations furthermore predict that the main external contribution to the build-up of the stellar halo came from the accretion of a few massive (i.e., $10^8$ to $10^9$ $M_\odot$) luminous satellites which merged with our galaxy more than 9 Gyr ago (for example, De Lucia & Helmi, 2008). This has recently been confirmed observationally by the discovery of the stellar debris of the Gaia-Enceladus galaxy, which merged with the Milky Way 10 Gyr ago, and then had a mass of $2.4 \times 10^9$ $M_\odot$ (Helmi et al., 2018). Low-mass satellite accretion continues until the present day, especially in the outer halo, although at a much reduced rate. These small satellites are minor contributors to the stellar mass of the halo.

The remnants of such accretion events are kinematically coherent substructures in the Galactic halo. Some of them also remain spatially coherent, so that they can be detected in wide-field imaging surveys, and many have indeed been discovered during the last decade (for example, Malhan et al., 2018 and references therein).

Existing spectroscopic samples suggest that at low metallicities, the α-elements, Fe-peak elements and neutron-capture elements in stars of classical dwarf spheroidal galaxies (dSphs) and Milky Way halo stars show very similar trends (for example, Mashonkina et al., 2017). In ultra-faint dwarf galaxies (UFDs), distinctly low abundances of neutron-capture elements and low [Sr/Ba] values (for example, Koch et al., 2013), and the abundance signature of a single *r*-process enrichment event (Ji et al., 2016) have been observed. Therefore, these abundance signatures can be used for associating kinematically identified groups of halo field stars with dSphs or UFDs. Furthermore, the location of the "knee" in the [α/Fe] vs. [Fe/H] abundance ratio diagram[a] (i.e., the value of [Fe/H] at which supernovae of type Ia start to contribute significantly to the chemical enrichment of the galaxy, thereby decreasing [α/Fe]) can be used to constrain the stellar mass of the host galaxy (for example Hendricks et al., 2014). With a large enough sample of halo stars it is therefore possible to identify the building blocks of the Galactic halo, determine their quantity and properties, and test numerical simulations of galaxy formation.

Additional significant contributions to the build-up of the Milky Way halo have been made by the accretion of stars from globular clusters (for example Martell et al., 2011). These stars can be identified in the halo field by their characteristic light element abundance ratios (Bastian & Lardo, 2018).

Our survey will also aim to significantly increase the sample of metal-poor stars (i.e., stars in which the abundances of elements heavier than helium are reduced by more than a factor of ten relative to the Sun) in the Galactic halo. These stars are tracers of the early chemical evolution of the Galaxy (Frebel & Norris, 2015). The abundances of the elements in their atmospheres provide us with information not only on the earliest phases of chemical enrichment of the Universe, but also on the nucleosynthetic processes contributing to the enrichment. In addition, they provide observational constraints on the physics of star formation processes in metal-poor environments and properties of the first generation of stars (for example, the initial mass function and rotation speeds).

## Specific scientific goals

The goals of the 4MOST Milky Way Halo High-Resolution Survey are:
– identification and determination of the elemental abundance patterns of stars that (a) formed *in situ* in the Galaxy, (b) were contributed from a few major accretion events, or (c) were contributed by minor, low-mass accretion events;
– identification of stars that were accreted to the halo from globular clusters, and quantification of their contribution to the build-up of the halo;
– studying the earliest phases of chemical evolution of the Milky Way and the nucleosynthetic processes involved by means of very metal-poor (i.e., [Fe/H] < −2.0) halo stars.

To reach these goals, we plan to determine the abundances of 10–20 elements in more than 1.5 million stars at high galactic latitude; the number of elements will depend on the stellar parameters, including [Fe/H]. By applying our selection criteria (Table 1) to the Gaia DR2 mock stellar catalogue of Rybizki et al. (2018), we estimate that our current catalogue contains about 200 000 genuine halo stars. This is the number of stars needed for the characterisation of the 300–600 kinematically coherent groups of stars that are expected to be present in the Galactic halo from cosmological simulations (for example, Gomez et al., 2013), because several hundred stars per group are needed to accurately sample the multi-dimensional abundance space. Note that most dSphs experienced extended star formation histories, so a wide range in [Fe/H] needs to be sampled.

We do not want to select these 200 000 genuine halo stars by kinematic criteria, to avoid kinematic biases in our sample. Furthermore, precise radial velocities cannot be acquired by the Gaia Radial Velocity Spectrometer for the majority of our stars, because most of them are too faint — Gaia will obtain radial velocities with a precision better than 1 km s$^{-1}$ only for G dwarf and K giant stars brighter than $G$ = 12.3 and 12.8 magnitudes (Vega), respectively. Therefore, the third component of their space motions will be known only *a posteriori*, once the 4MOST spectra have been obtained.

| Criterion # | Bright survey | Faint survey | Deep survey |
|---|---|---|---|
| 1 | +20° ≥ dec ≥ −80° | | Selected areas |
| 2 | \|b\| > 20° | | |
| 3 | [Fe/H] < −0.5 | | |
| 4 | 12.0 ≤ G ≤ 14.5 | 14.5 < G ≤ 15.5 | 15.5 < G ≤ 17.0 |
| 5 | 0.15 ≤ $(G_{BP}-G_{RP})_0$ ≤ 1.10 | | |
| 6 | (1.10 < $(G_{BP}-G_{RP})_0$ ≤ 1.60) & ($M_G$ < 3.5) | | |
| Total number of targets | 1 150 000 | 800 000 | 26 000 |
| Targets at [Fe/H] < −2.0 | 13 000 | 18 000 | 100 |

Table 1. Target selection criteria for the three sub-surveys. The criteria 5 and 6 select dwarf/sub-giant, and giant stars, respectively. The logical combination of the criteria is "1 and 2 and 3 and 4 and (5 or 6)".

At the low-metallicity end, we will be able to increase the sample of halo stars with elemental abundances based on high-resolution (i.e., $R = \lambda/\Delta\lambda$ > 18 000) spectroscopy by almost two orders of magnitude. At the time of writing, the Stellar Abundances for Galactic Archaeology (SAGA)[1] database lists 323 such stars at [Fe/H] < −2.0 with available high-resolution spectra, while our target catalogue contains 31 000 stars in that metallicity range, of which we expect ~ 24 000 to be observed.

Extrapolating from the currently existing samples, we estimate that we will find ~ 200 new stars at [Fe/H] < −4.0, compared to the 24 that are known today, according to the SAGA database. These stars are presumably second-generation stars; i.e., their elemental abundance patterns are the imprints of the supernova explosions of the first stars in the Universe. Therefore, we will be able to derive indirectly properties of the first generation of stars (including, for example, their mass distribution), and to infer information on the physics of star formation in low-metallicity environments. Furthermore, the larger sample will give us a chance of identifying objects that are too rare to be included in the currently existing samples.

## Science requirements

Since the density of stars bright enough for high-resolution spectroscopy with a 4-metre telescope is low in fields at high galactic latitude, this survey needs large sky coverage. We are therefore aiming at a survey area of 14 000 square degrees.

At an average target density of about 450 dwarf and giant stars per 4.2 square degrees of the 4MOST field down to $G$ = 15.5 magnitudes (but varying strongly with galactic latitude), this will lead to a sample of more than 1.5 million stars. Our current target catalogue covers 18 700 square degrees (see Figure 2), including a total of about 4 700 square degrees at declinations, dec > +5 degrees or dec < −70 degrees (i.e., outside the fiducial survey footprint, see Guiglion et al., p. 17). However, we expect that only a small fraction of these fields outside the fiducial footprint can be observed, because of unfavourable observing conditions (for example, higher airmass or prevailing northern winds at Paranal), and the total amount of observing time available in a five-year survey.

For characterising the abundance patterns, all the relevant element groups need to be covered, including the light elements (for example, C), α-elements (for example, Mg, Ca), and neutron-capture elements (for example, Ba, Sr, Eu). In the optical spectra of very metal-poor stars, there are very few absorption lines at λ > 4500 Å (see Hansen et al., 2015 for a detailed study). For these stars, we will mostly rely on the spectra acquired with the blue arm of the 4MOST high-resolution spectrograph. However, valuable additional information can be derived, for example, from the Mg I b triplet lines covered by the green arm, and from the Hα and the Li I 6707 Å lines covered by the red arm. Furthermore, the green and red arm spectra of higher metallicity (i.e., −2.0 < [Fe/H] < −0.5) stars will contain many more detectable lines, which will increase the precision of the derived elemental abundances.

Elemental abundance ratios with a precision of $\sigma_{[X/Y]}$ < 0.2 dex are typically needed to distinguish between different stellar populations, and to determine the potential origin of the stars. For example, stars with [α/Fe] ~ 0.0, which are characteristic of dwarf galaxies (for example, Tolstoy et al., 2009) — as opposed to the canonical [α/Fe] = +0.4 in the Milky Way halo — can be recognised reliably only at this precision in abundance ratio measurement. The [α/Fe] ratios of the inner and outer halo are separated by only 0.1 dex (Nissen & Schuster, 2015), but several α-elements can be combined to increase the precision.

Information on the binary status of the targets of our survey is important for the interpretation of their elemental abundance patterns, since it provides constraints on the nucleosynthetic origin of, for example, the neutron-capture elements in the atmospheres of the observed stars. Therefore, we want as many targets as possible to be re-observed on timescales of months to years, so that radial velocity variations can be detected.

### Target selection and survey area

The targets for our survey will be selected from the Gaia Data Release 3 catalogue, based on their apparent Gaia magnitude ($G$), de-reddened $G_{BP}$–$G_{RP}$ colour (i.e., $(G_{BP}$–$G_{RP})_0$), where $G_{BP}$ and $G_{RP}$ are the integrated Gaia Blue and Red Photometer magnitudes, respectively, and absolute $G$ magnitude ($M_G$), determined using Gaia parallaxes (or upper limits). For the metallicity selection, we will use the best data available at the time the 4MOST survey starts; for example, from the SkyMapper survey (Casagrande et al., 2019), Gaia BP spectra, or at dec > 0 degrees from the Pristine survey (Starkenburg et al., 2017). The selection criteria are summarised in Table 1.

Since we are targeting halo stars, we have restricted the survey footprint to galactic latitudes of $|b|$ > 20 degrees, taking advantage of the fact that the high-resolution fibres of 4MOST will not be used by the extragalactic surveys. In the range 20 < $|b|$ < 30 degrees, there will be a significant overlap with the targets of 4MIDABLE-HR (Bensby et al., p. 35). The overlap is currently estimated to be about 440 000 stars, so a joint target catalogue will be created and observed spectra will be exchanged.

| Star type | S/N per pixel[b] in survey | | |
|---|---|---|---|
| | Bright | Faint | Deep |
| Dwarf & subgiant | 50 | 25 | 25 |
| Giant | 30 | 15 | 15 |

Table 2. Spectral success criteria. The S/N is measured in the wavelength region 4261–4270 Å, which is free of strong spectral lines in the stellar parameter range covered by our survey. The criteria reflect the different science goals of the sub-surveys, as well as the different line strengths in dwarf and giant stars of a given metallicity.

Our metallicity cut ([Fe/H] < –0.5) will ensure that our survey will include the interesting metallicity range –1.5 < [Fe/H] < –0.5, over which the [α/Fe] abundance ratios allow us to discriminate between the inner and outer halo populations (Nissen & Schuster, 2010). On the other hand, the criterion removes stars of the disc populations in the Milky Way.

For harmonisation with the maximum total exposure times per high-galactic-latitude field needed by the low-resolution surveys, we have defined three sub-surveys: (1) a bright survey of stars in the range 12 ≤ $G$ ≤ 14.5 magnitudes, allowing us to acquire spectra with a signal-to-noise ratio (S/N) larger than 50 per pixel[b] in the continuum at 4260 Å with 4MOST in less than 2 hours; (2) a faint survey (14.5 < $G$ ≤ 15.5 magnitudes); and (3) a deep survey (15.5 < $G$ ≤ 17.0 magnitudes). In the latter two sub-surveys, the S/N requirements are reduced by a factor of two with respect to the bright survey. The area of the deep survey will be aligned with the sky regions in which other 4MOST surveys will obtain longer exposures, for example, the WAVES-Wide and WAVES-Deep surveys. The fainter magnitude limit in these fields will allow us to observe giant stars at distances of up to 25 kpc, so the fraction of outer halo stars among our targets is expected to be significantly higher in these fields.

We note that the magnitude ranges of the 4MOST high- and low-resolution surveys of the Milky Way halo are complementary. In the former, the brighter magnitude limit will result in a sample dominated by the inner halo population, while the latter will focus on exploring the outer halo.

All three of our sub-surveys target dwarf, subgiant, and giant stars. The blue colour limit of $(G_{BP}$–$G_{RP})_0$ = 0.15 magnitudes is chosen to match the colours of 13-Gyr-old ultra-metal-poor (i.e., [Fe/H] < –4.0) stars near the main-sequence turnoff; the red limit of $(G_{BP}$–$G_{RP})_0$ = 1.6 magnitudes ensures that metal-poor K giant stars are included, while main-sequence stars of spectral type M or later are removed. To remove foreground K dwarf stars belonging to the Galactic disc populations, we have added an absolute magnitude criterion for the stars at $(G_{BP}$–$G_{RP})_0$ > 1.1 magnitudes. With that colour limit, metal-poor G dwarf stars are still included. The absolute magnitude criterion for giant stars is $M_G$ < 3.5 magnitudes. A colour-magnitude diagram of the targets in a narrow range around [Fe/H] = –1.0 is shown in Figure 1.

### Spectral success criteria and figure of merit

Our spectral success criteria are chosen such that precise (i.e., $\sigma_{[X/Y]}$ < 0.1–0.2 dex) elemental abundances of up to 20 elements can be determined for the targets of the bright sub-survey, while for the fainter targets at least a rough characterisation of the abundance patterns of the stars will be possible by means of abundances of the most important elements with a typical precision of $\sigma_{[X/Y]}$ < 0.2–0.3 dex, depending on the elements and the stellar parameters. This will make it possible to, for example, identify metal-poor stars enhanced in carbon, neutron-capture elements, or combinations thereof. Example spectra are shown in Figure 3, and the criteria are listed in Table 2.

The current definition of the figure of merit (FoM) of this survey increases linearly with the number of successfully observed targets, and it is 1.0 if 1.5 million stars have been observed successfully. However, we are considering implementing a numerical scheme that takes into account "partially successful" observations such as that outlined in the White Paper of the 4MOST Consortium Milky Way Halo Low-Resolution Survey (see Helmi et al., p. 23), for the reasons discussed there.

Figure 1. Colour-magnitude diagram of the targets of the three sub-surveys. We show stars at [Fe/H] = –1.0 ± 0.1, taken from the Gaia DR2 mock stellar catalogue (Rybizki et al., 2018), and compare them with isochrones for [Fe/H] = –1.0 and ages 10, 12, and 13 Gyr. The numbers inside the panels are the distance ranges covered by the targets at the given absolute magnitude. The absence of stars on the subgiant branch (i.e., roughly in the absolute magnitude range $3 < M_G < 4$ magnitudes) is an artefact introduced by selecting only stars for which [Fe/H] has been determined with a precision of better than 0.3 dex. Note that the $(G_{BP}-G_{RP})_0$ colours of main-sequence turnoff stars at [Fe/H] = –4.0 are considerably bluer than those of the stars shown here, hence our choice of a colour cut-off at $(G_{BP}-G_{RP})_0 = 0.15$ magnitudes.

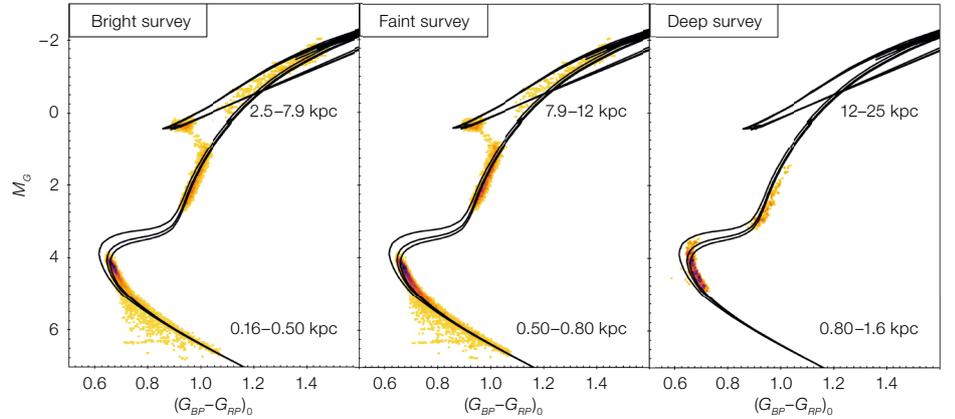

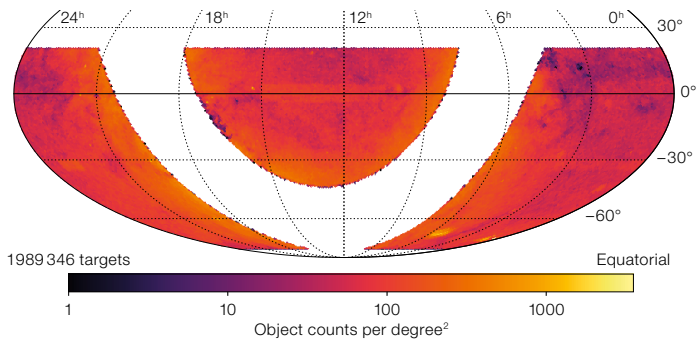

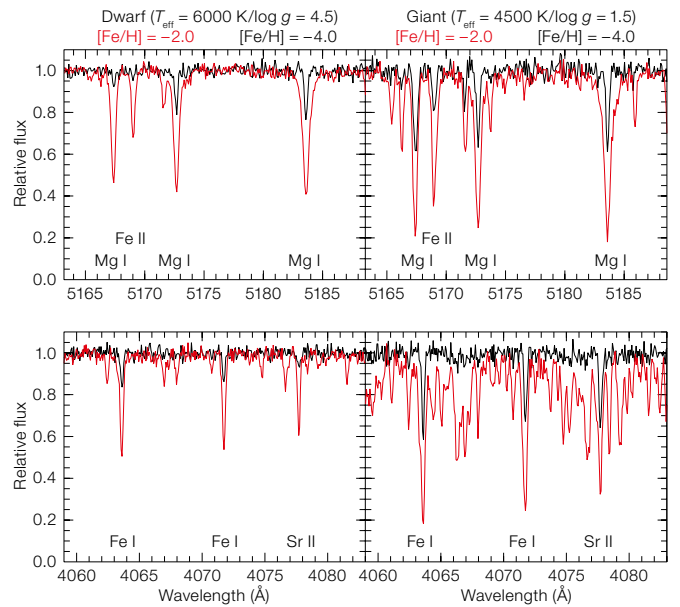

Figure 2. Combined target density of the three sub-surveys. The areas covered by our deep survey can be recognised by their slightly higher target densities compared to the combined target density of the other two sub-surveys, for example, at dec ~ 0 degrees and 10 < right ascension < 15 hours, which is one of the WAVES-Wide fields (see WAVES White Paper of Driver et al., p. 46). Note that targets with dec > +5 and dec < –70 degrees have a smaller likelihood of being observed in the baseline survey strategy (see Guiglion et al., p. 17).

Figure 3. Synthetic spectra of a dwarf star (left panels) and a giant star (right panels) at the S/N required to reach the science goals of the bright survey (see Table 1). At that S/N, several Fe lines, as well as the Mg $b$ triplet lines, are clearly detected even at [Fe/H] = –4.0.

Acknowledgements

This work is supported by Sonderforschungsbereich SFB 881 "The Milky Way System" (subprojects A3, A4 and A9) of the German Research Foundation (DFG), by the German Federal Ministry of Education and Research (BMBF) through Verbundforschungsprojekt 05A17VH4, by the Observatoire de Paris, and the project grant "The New Milky Way" from the Knut and Alice Wallenberg Foundation.

Links

[1] Stellar Abundances for Galactic Archaeology Database: http://sagadatabase.jp/

Notes

[a] [X/Y] = $\log_{10}(N_X/N_Y)_\star - \log_{10}(N_X/N_Y)_\odot$, where $N_X$ and $N_Y$ are the number densities of the elements X and Y, respectively.
[b] For a conversion from S/N per pixel into S/N Å$^{-1}$, a factor of 3.3 has to be applied (see 4MOST overview paper by de Jong et al., p. 3).